\documentclass[12pt]{llncs}
\usepackage{fullpage,graphicx}

\newcommand{\ignore}[1]{}
\newcommand{\Oh}[1]
    {\ensuremath{\mathcal{O}\!\left({#1}\right)}}
\newcommand{\polylog}
    {\ensuremath{\mathrm{polylog}}}
\newcommand{\str}
    {\ensuremath{\mathrm{str}}}

\begin{document}

\title{Linear-Space Substring Range Counting\\over Polylogarithmic Alphabets}
\author{Travis Gagie\inst{1} \and Pawe{\l} Gawrychowski\inst{2}}
\institute{Aalto University, Finland\\
\email{travis.gagie@aalto.fi}\\[1ex]
\and Max Planck Institute, Germany\\
\email{gawry@cs.uni.wroc.pl}}
\maketitle

\begin{abstract}
Bille and G{\o}rtz (2011) recently introduced the problem of substring range counting, for which we are asked to store compactly a string $S$ of $n$ characters with integer labels in \([0, u]\), such that later, given an interval \([a, b]\) and a pattern $P$ of length $m$, we can quickly count the occurrences of $P$ whose first characters' labels are in \([a, b]\).  They showed how to store $S$ in $\Oh{n \log n / \log \log n}$ space and answer queries in $\Oh{m + \log \log u}$ time.  We show that, if $S$ is over an alphabet of size \(\polylog (n)\), then we can achieve optimal linear space.  Moreover, if \(u = n\,\polylog (n)\), then we can also reduce the time to $\Oh{m}$.  Our results give linear space and time bounds for position-restricted substring counting and the counting versions of indexing substrings with intervals, indexing substrings with gaps and aligned pattern matching.
\end{abstract}

\section{Introduction} \label{sec:introduction}

Bille and G{\o}rtz~\cite{BG11a} recently introduced the problem of substring range reporting, for which we are asked to store compactly a string $S$ of $n$ characters with integer labels in \([0, u]\), such that later, given an interval \([a, b]\) and a pattern $P$ of length $m$, we can quickly report the occurrences of $P$  whose first characters' labels are in \([a, b]\).  They showed how to store $S$ in $\Oh{n \log^{\epsilon} n}$ space on a word RAM and answer queries in $\Oh{m + t}$ time, where $\epsilon$ is an arbitrary positive constant and $t$ is the number of occurrences reported.  We work in the word RAM model as well so, unless otherwise specified, throughout this paper we measure space in words.  They also showed that this gives the same space and time bounds for position-restricted substring search~\cite{MN07}, indexing substrings with intervals~\cite{CIKRW10} and indexing substrings with gaps~\cite{IR09}.  Their solution consists of a suffix tree, a data structure for 2-dimensional range reporting and several instances of a data structure for 1-dimensional range reporting.  Calculation shows that, if we replace the 1-dimensional instances by bitvectors, then they take only $\Oh{n}$ space.  This does not improve the overall bound, however, because the 2-dimensional data structure still takes $\Oh{n \log^{\epsilon} n}$ space.  Chien, Hon, Shah and Vitter~\cite{CHSV08} proved that, in the weaker pointer-machine model, any solution for position-restricted substring search with $\Oh{m \polylog (n) + t}$ query time must use \(\Omega (n \log n / (\log \log n)^2)\) space, even when the alphabet has constant size.  By Bille and G{\o}rtz' reduction, this lower bound holds for substring range search as well, even when the alphabet has constant size and \(u = n\).

In an updated version of their paper~\cite{BG11b}, Bille and G{\o}rtz introduced the related problems of substring range counting, for which we are asked only to count the occurrences of $P$  whose first characters' labels are in \([a, b]\), and substring range emptiness, for which we are asked whether there exist any such occurrences.  For the counting problem, we could restrict our attention to intervals of the form \([1, b]\), but we consider general intervals for consistency with Bille and G{\o}rtz' paper.  They gave a solution to the counting problem that uses $\Oh{n \log n / \log \log n}$ space and $\Oh{m + \log \log u}$ query time, and a solution to the emptiness problem that uses $\Oh{n \log \log (n + u)}$ space and $\Oh{m}$ query time.  (Although they claimed to use $\Oh{n \log \log u}$ space for the emptiness problem, this seems to be under the assumption \(u \geq n\).)  They noted that solutions to the substring range counting and emptiness problems give the same space and time bounds for position-restricted substring counting and emptiness and the counting and emptiness versions of indexing substrings with intervals and indexing substrings with gaps.  As before, each solution consists of a suffix tree, one data structure for 2-dimensional range queries and several instances of a data structure for 1-dimensional range queries.  For range counting and range emptiness, however, there are fast linear-space solutions known, leaving open the possibility that simple modifications of their solutions give better bounds.  Notice that, unless we restrict the labelling, linear space is optimal when \(\log u = \Omega (\log n)\).  It makes no difference here, but by $\log$ we always mean $\log_2$.

In this paper we show that, if $S$ is over a polylogarithmic alphabet (i.e., of size \(\polylog (n) = \log^{\mathcal{O} (1)} n\)) then we can reduce the space bound for substring range counting to $\Oh{n}$ while still answering queries in $\Oh{m + \log \log u}$ time.  If \(u = n\,\polylog (n)\) --- as in position-restricted substring counting, indexing with intervals and indexing with gaps --- then we can also reduce the time bound to $\Oh{m}$.  (Notice linear time is unfortunately not necessarily optimal, as we need only \(\Oh{m \log \log n / w}\) time to read $P$, where \(w = \Omega (\log (n + u))\) is the word size.)  In this case, our bounds for substring range counting are strictly better than Bille and G{\o}rtz' bounds even for substring range emptiness, which is a special case of counting.  By Bille and G{\o}rtz' arguments, our results give linear space and time bounds for position-restricted substring counting and the counting versions of indexing substrings with intervals and indexing substrings with gaps.  We show they also imply linear space and time bounds for the counting version of aligned pattern matching~\cite{Tha11}.

It is not difficult to see that we generally cannot use both linear space and linear time when $u$ is unrestricted: suppose that, given $n$ elements from a universe of size $u$, we store them in an array and assign them as labels to the characters of a unary string; we can implement rank queries as substring range counting queries for patterns of length 1, and select queries as array accesses; it follows that, if we could answer substring range counting queries in $\Oh{m}$ time, then we could answer predecessor queries in $\Oh{1}$ time.  On the other hand, although our own approach seemingly will not work, we see no obvious reason why it is impossible to use both linear space and linear time when, say, both $u$ and the alphabet size are equal to $n$.  We leave this as an open problem.

\section{Applications} \label{sec:applications}

For position-restricted substring counting, we are asked to store $S$ compactly such that, given a pattern $P$ and an interval \([i, j]\) with \(1 \leq i \leq j \leq n\), we can quickly count the occurrences of $P$ starting in \(S [i, j]\).  As Bille and G{\o}rtz noted, to solve this problem via substring range counting, we simply assign each character of $S$ a label equal to its position in $S$.  For the counting version of the slightly more complicated problem of indexing substrings with intervals, we are asked to store $S$ and a set $\pi$ of intervals in \([1, n]\) such that, given $P$ and \([i, j]\), we can quickly count the occurrences of $P$ that start in \([i, j]\) {\em and} in one of the intervals in $\pi$.  Bille and G{\o}rtz noted that, to solve this problem, we change to 0 the labels of all characters not in any interval in $\pi$.

For the counting version of index substrings with gaps, we are given an integer $d$ and asked to store $S$ compactly such that, given two patterns $P_1$ and $P_2$ of total length $m$, we can quickly count the positions in $S$ where there are occurrences of $P_1$ followed by $d$ characters followed by occurrences of $P_2$.  Solving this problem via substring range counting was Bille and G{\o}rtz' most sophisticated reduction.  We sort the reversed prefixes of $S$ into lexicographic order and assign the rank of \(S [1..i]\) reversed as the label of \(S [i + d + 1]\).  Given $P_1$ and $P_2$, we compute the interval containing the lexicographic ranks of reversed prefixes that start with $P_1$ reversed, then count the occurrences of $P_2$ whose first characters' labels are in that interval.

For aligned pattern matching, we are given two strings $S_1$ and $S_2$ of total length $n$ and asked to store them compactly such that later, given two patterns $P_1$ and $P_2$ of total length $m$, we can quickly find all the locations where $P_1$ occurs in $S_1$ and $P_2$ occurs in $S_2$ in the same position.  Thankachan~\cite{Tha11} noted that this problem can be solved directly via 2-dimensional range reporting, using $\Oh{n \log^\epsilon n}$ space and $\Oh{\log \log n + t}$ time, where $t$ is the number of such locations.  He then showed how to store $S_1$ and $S_2$ in compressed form, but this solution takes $\Oh{m + \log^{4 + \epsilon} n + t}$ time when the lengths of $P_1$ and $P_2$ are both in \(\Omega (\log^{2 + \epsilon} n)\), and $\Oh{m + \sqrt{n t} \log^{2 + \epsilon} n}$ time otherwise.

It is straightforward to reduce aligned pattern matching to substring range reporting: we simply assign each character \(S_1 [i]\) in $S_1$ the lexicographic rank of the $i$th suffix of $S_2$ (i.e., the sequence of labels is the suffix array of $S_2$).  To answer a query, we first find the interval containing the lexicographic ranks of the suffixes of $S_2$ that start with $P_2$.  We then search for occurrences of $P_1$ in $S_1$ whose first characters' labels are in that interval.  This reduction also works for the counting versions of these problems.  More generally, we might be given a library of strings and a function $f$ mapping positions to positions, and asked to count the times $P_1$ occurs in a position $i$ in the library and $P_2$ occurs in position \(f (i)\).  In this general case $\Oh{n}$ space is optimal, even just to store the labels.

Since we use only labels in \([1, n]\) in any of these reductions, it follows by our bounds for substring range counting that if $S$ is over a polylogarithmic alphabet, then we can solve all these counting problems using linear space and time, i.e., $\Oh{n}$ space and $\Oh{m}$ time.

Finally, we note that Bille and G{\o}rtz' result can be modified to very slightly speed up some grammar-based self-indexes.  For example, Gagie, Gawrychowski, K\"arkk\"ainen, Nekrich and Puglisi~\cite{GGKNP12} recently showed how, given a balanced straight-line program for a string $S$ of length $n$ whose LZ77 parse consists of $z$ phrases, we can add $\Oh{z \log \log z}$ words such that later, given a pattern $P$ of length $m$, we can find all $t$ occurrences of $P$ in $S$ in $\Oh{m^2 + (m + t) \log \log n}$ time.  Following Kreft and Navarro~\cite{KN11} and previous authors, they use two Patricia trees~\cite{Mor68} and a data structure for 2-dimensional range reporting to find the occurrences of $P$ that cross phrase boundaries in the parse.  Without going into too much detail, they split $P$ into a prefix and suffix at every positive position, then search for the reversed prefix in one Patricia tree and the suffix in the other, then use range reporting to determine which phrases are preceded by the the prefix and followed by the suffix.  If we store data structures for 1-dimensional range reporting (or bitvectors) at the uppermost \(\log \log n\) levels of one of the Patricia trees --- much like Bille and G{\o}rtz do in one suffix tree for indexing substrings with gaps --- then the time bound for searching shrinks (albeit very slightly) to $\Oh{m^2 + t \log \log n}$, while the space bound is not affected.  If $S$ is over a polylogarithmic alphabet and we are interested only in determining whether $P$ occurs in $S$ at all, but not where nor how often, then we can use our results from this paper to reduce the time bound further, to $\Oh{m^2}$, while simultaneously reducing the added space to $\Oh{z}$ words.  We will give full details in the full version of that paper.

\begin{theorem} \label{thm:grammars}
Given a balanced straight-line program for a string $S$ of length $n$ whose LZ77 parse consists of $z$ phrases, we can all $\Oh{z \log \log z}$ words such that, given a pattern $P$ of length $m$, we can find all $t$ occurrences of $P$ in $S$ in $\Oh{m^2 + t \log \log n}$ time.
\end{theorem}

\section{Preliminaries} \label{sec:preliminaries}

The suffix tree for $S$ is the compacted trie storing the suffixes of $S$, so each edge is associated with a substring of $S$, called its label.  A child query at a node $v$ takes a character $c$ and returns a pointer to the unique child of $v$ that is reached by an edge whose label starts with $c$.  The concatenation of edge labels on the path from the root to $v$ is denoted \(\str_S (v)\).  We say $v$'s string depth is \(|\str_S (v)|\).  An interval query at $v$ returns the interval of lexicographic ranks of the suffixes of $S$ that start with \(\str_S (v)\).  If $v$ is the shallowest node such that $P$ is a prefix of \(\str_S (v)\), then we call $v$ the locus of $P$.  We use suffix trees in essentially the same way as Bille and G{\o}rtz do, so we refer the reader to their paper for more discussion.  The key fact for us is that, using perfect hashing~\cite{FKS84}, we can store the suffix tree for $S$ in $\Oh{n}$ space such that child and interval queries take $\Oh{1}$ time.

The 2-dimensional range counting problem is to store compactly a set of points in the plane such that, given a query rectangle, we can quickly count the points it contains.  Following Bille and G{\o}rtz, we use the following theorem; as they noted, combined na\"{i}vely with a suffix tree, this result can be used as an $\Oh{n}$-space, $\Oh{m + \log n / \log \log n + \log \log u}$-time solution for substring range counting.

\begin{theorem}[J\'aJ\'a, Mortensen and Shi~\cite{JMS04}] \label{thm:JMS04}
We can store $n$ 2-dimensional points in $\Oh{n}$ space such that range counting queries take $\Oh{\log n / \log \log n}$ time.
\end{theorem}

\noindent
We use the next theorem simply to map the interval \([a, b]\) to the subinterval of labels in \([a, b]\) that actually occur in $S$.  P\u{a}tra\c{s}cu~\cite{Pat08} showed that, if \(u = n\,\polylog (n)\), then we can store a bitvector with any redundancy in \(n / \polylog (n)\) that supports rank and select (and, thus, predecessor) queries in $\Oh{1}$ time.

\begin{theorem}[Willard~\cite{Wil83} and P\u{a}tra\c{s}cu~\cite{Pat08}] \label{thm:predecessor}
We can store $n$ integers from \([0, u]\) in $\Oh{n}$ space such that predecessor queries take $\Oh{1}$ time if \(u = n\,\polylog (n)\) and $\Oh{\log \log u}$ time otherwise.
\end{theorem}

\noindent
A rank query on $S$ takes as arguments a character $c$ and a position $i$ and returns the number of occurrences of $c$ in the prefix of $S$ of length $i$.  A select query takes $c$ and a rank $j$ and returns the position of the $j$th occurrence of $c$ in $S$.  Ferragina, Manzini, M\"akinen and Navarro~\cite{FMMN07} showed how to store $S$ in compressed space such that these queries take $\Oh{1 + \frac{\log \sigma}{\log \log n}}$ time, where $\sigma$ is the alphabet size; this time bound is $\Oh{1}$ when \(\sigma = \polylog (n)\).

\begin{theorem}[Ferragina, Manzini, M\"akinen and Navarro~\cite{FMMN07}] \label{thm:FMMN07}
We can store a string of $n$ characters from an alphabet of size \(\polylog (n)\) in $\Oh{n \log \sigma}$ bits such that rank and select queries take $\Oh{1}$ time.
\end{theorem}

\section{Data Structure} \label{sec:data structure}

Following Bille and G{\o}rtz, we build a suffix tree and divide it into a top tree, consisting of all nodes whose string depths are at most \(\log n / \log \log n\), and a forest of bottom trees, induced by the remaining nodes.  We build a data structure for 2-dimensional range counting according to Theorem~\ref{thm:JMS04}, storing a point \((x, y)\) if and only if the first character in the lexicographically $x$th suffix is labelled $y$.  If \(m > \log n / \log \log n\) we use the suffix tree to find the interval containing the lexicographic ranks of the suffixes starting with $P$, then we count the points in the product of that interval and \([a, b]\), all in $\Oh{m}$ time.  Our solution differs in how we deal with the case when \(m \leq \log n / \log \log n\).

We sort the characters in $S$ by their labels and store the resulting string $S_r$ according to Theorem~\ref{thm:FMMN07} at the root $r$ of the suffix tree.  Suppose a node $v$ and one of its children $v'$ both have string depth less than \(\log n / \log \log n\); the edge from $v$ to $v'$ is labelled \(c \circ \alpha\), where $\circ$ denotes concatenation; and we store the string $S_v$ at $v$.  Then we build a new string $S_{v'}$ from $S_v$ and store it at $v'$.  To build $S_{v'}$, we discard all the characters of $S_v$ not equal to $c$ and replace each occurrence \(S [i]\) of $c$ in $S_v$ by \(S [i + |\alpha| + 1]\) (or \$ if \(i + |\alpha| + 1 = n + 1\)).  We store all the strings at each depth in the suffix tree according to Theorem~\ref{thm:FMMN07}, which takes a total of $\Oh{(n \log \sigma) (\log n / \log \log n)}$ bits, where $\sigma$ is again the alphabet size.  Assuming \(\sigma = \polylog (n)\), all the strings take a total of $\Oh{n}$ space and rank queries take $\Oh{1}$ time.

\begin{figure}[t!]
\begin{centering}
\includegraphics[width = 75ex]{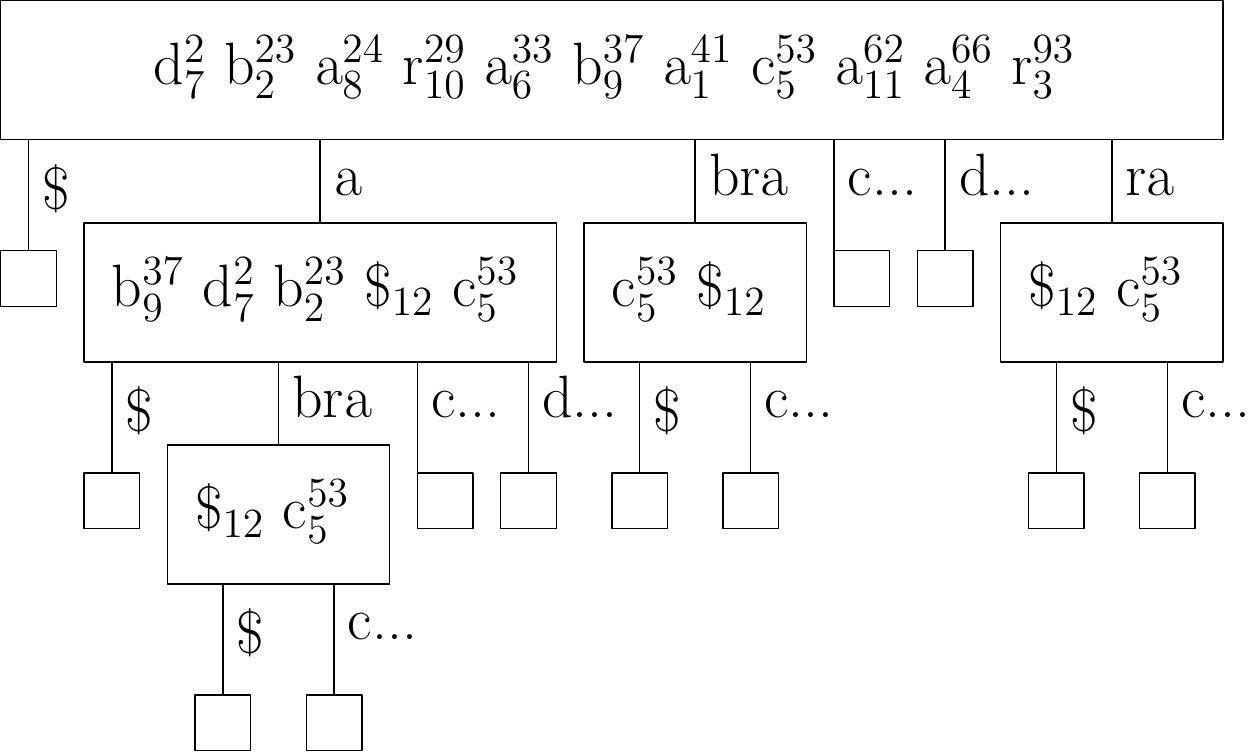}
\caption{The strings we store at the suffix tree's nodes when \(S = \mathrm{a_1^{41}\ b_2^{23}\ r_3^{93}\ a_4^{66}\ c_5^{53}\ a_6^{33}\ d_7^{2}\ a_8^{24}\ b_9^{37}\ r_{10}^{29}\ a_{11}^{62}}\).}
\label{fig:suffix tree}
\end{centering}
\end{figure}

Figure~\ref{fig:suffix tree} shows the strings we store at the suffix tree's nodes when \(S = \mathrm{a_1^{41}\ b_2^{23}\ r_3^{93}\ a_4^{66}\ c_5^{53}}\) \(\mathrm{a_6^{33}\ d_7^{2}\ a_8^{24}\ b_9^{37}\ r_{10}^{29}\ a_{11}^{62}}\), where characters' positions in $S$ are shown as subscripts and their labels (pseudo-randomly chosen) are shown as superscripts.  We have written the positions and labels in the figure to give insight into the construction, even though we do not actually store them.  The string at the root consists of the characters of $S$ sorted by label.  To build the string at the root's third child, for example, we look at the edge from the root to that child and see its label begins with `b' and has length 3; we find all the occurrences of `b' in the root's string --- i.e., \(\mathrm{b_2^{23}}\) and \(\mathrm{b_9^{37}}\) --- and replace them by the characters 3 positions later in $S$ --- i.e., \(\mathrm{c_5^{53}}\) and \(\mathrm{\$_{12}}\).

We build a predecessor data structure according to Theorem~\ref{thm:predecessor} and use it to store the characters' labels in $S$.  We also store a partial-sum data structure for these labels' frequencies.  Together, these data structures take $\Oh{n}$ space and allow us to compute the interval in $S_r$ containing characters with labels in \([a, b]\), in $\Oh{1}$ time if \(u = n\,\polylog (n)\) and in $\Oh{\log \log u}$ time otherwise.

By induction, the characters in $S_v$ are the ones immediately following occurrences of \(\str_S (v)\) in $S$; they are sorted by the labels of the first characters of those occurrences of \(\str_S (v)\).  Suppose we know the interval in $S_v$ containing characters immediately following in $S$ occurrences of \(\str_S (v)\) whose first characters' labels are in \([a, b]\).  Notice that the ranks of the first and last occurrences of $c$ in that interval of $S_v$, which we can find in $\Oh{1}$ time, are the endpoints of the interval in $S_{v'}$ containing characters immediately following in $S$ occurrences of \(\str_S (v') = \str_S (v) \circ c \circ \alpha\) whose first characters' labels are in \([a, b]\).  Since the length of this interval in $S_{v'}$ is the number of such occurrences of \(\str_S (v)\), we can also count them in $\Oh{1}$ time.

If \(m \leq \log n / \log \log n\) then we descend from the root of the suffix tree to the locus of $P$, at each node $v$ computing the interval in $S_v$ containing characters immediately following in $S$ occurrences of \(\str_S (v)\) whose first characters' labels are in \([a, b]\).  When we reach the locus of $P$, we return the length of the interval.  This takes a total of $\Oh{m}$ time.

Suppose \(S = \mathrm{a_1^{41}\ b_2^{23}\ r_3^{93}\ a_4^{66}\ c_5^{53}\ a_6^{33}\ d_7^{2}\ a_8^{24}\ b_9^{37}\ r_{10}^{29}\ a_{11}^{62}}\) and we want to count the occurrences of \(P = \mathrm{ab}\) whose first characters' labels are in \([20, 40]\).  Using the predecessor and partial-sum data structure, we compute the interval \([2, 6]\) in the string stored at the root in Figure~\ref{fig:suffix tree}, that contains the characters with labels in \([20, 40]\).  With two rank queries, we determine that this interval contains the string's first and second occurrences of `a'.  We descend along the edge labelled `a' and consider the interval \([1, 2]\) in the string stored at the child.  With two more rank queries we determine there is only 1 `b' in that interval.  Therefore, there is only one such occurrence of $P$ in $S$.

\begin{theorem} \label{thm:range counting}
Suppose we are given a string $S$ of $n$ characters with integer labels in \([0, u]\), over an alphabet of size \(\polylog (n)\).  We can store $S$ in $\Oh{n}$ space such that, given an interval \([a, b]\) and pattern $P$ of length $m$, we can count the occurrences of $P$ whose first characters' labels are in \([a, b]\) using $\Oh{m}$ time if \(u = n\,\polylog (n)\) and $\Oh{m + \log \log u}$ time otherwise.
\end{theorem}

We note as an aside that, climbing back up to the root and using a select query at each step, we can find the position in the string stored at the root of the first character in an occurrence of $P$ whose label is in \([a, b]\), again in $\Oh{m}$ time.  We can store in $\Oh{n}$ space the permutation that maps characters in $S_r$ back to their positions in $S$, so we can return an example occurrence.  For the case when \(m > \log n / \log \log n\), we store an $\Oh{n}$-space data structure for 2-dimensional range reporting~\cite{CLP11} and use it to find a single example point; with the permutation, we then map that point back to a position in $S$.

As we pointed out in Section~\ref{sec:introduction}, our solution for substring range counting immediately gives the same space and time bounds for substring range emptiness.  By the reductions in Section~\ref{sec:applications}, we also have the following theorem.

\begin{theorem} \label{thm:applications}
For strings over polylogarithmic alphabets, we can solve position-restricted substring counting and the counting versions of indexing with intervals, indexing with gaps and aligned pattern matching, all using space linear in the string length and query time linear in the pattern length.
\end{theorem}

\subsection*{Acknowledgments}

Many thanks to Veli M\"akinen, Gonzalo Navarro, Simon Puglisi and Sharma Thankachan, for helpful discussions.

\bibliographystyle{plain}
\bibliography{counting}

\begin{thebibliography}{10}

\bibitem{BG11a}
P.~Bille and I.~L. G{\o}rtz.
\newblock Substring range reporting.
\newblock In {\em Proceedings of the 22nd Symposium on Combinatorial Pattern
  Matching (CPM)}, pages 299--308, 2011.

\bibitem{BG11b}
P.~Bille and I.~L. G{\o}rtz.
\newblock Substring range reporting.
\newblock Technical Report 1108.3683, {\tt www.arxiv.org}, 2011.

\bibitem{CLP11}
T.~M. Chan, K.~G. Larsen, and M.~P\u{a}tra\c{s}cu.
\newblock Orthogonal range searching on the {RAM}, revisited.
\newblock In {\em Proceedings of the 27th Symposium on Computational Geometry
  (SoCG)}, pages 1--10, 2011.

\bibitem{CHSV08}
Y.~F. Chien, W.-K. Hon, R.~Shan, and J.~S. Vitter.
\newblock Geometric {Burrows-Wheeler Transform}: Linking range searching and
  text indexing.
\newblock In {\em Proceedings of the Data Compression Conference (DCC)}, pages
  252--261, 2008.

\bibitem{CIKRW10}
M.~Crochemore, C.~S. Iliopoulos, M.~Kubica, M.~S. Rahman, , and T.~Walen.
\newblock Finding patterns in given intervals.
\newblock {\em Fundamenta Informaticae}, 101(3):173--186, 2010.

\bibitem{FMMN07}
P.~Ferragina, G.~Manzini, V.~M{\"a}kinen, and G.~Navarro.
\newblock Compressed representations of sequences and full-text indexes.
\newblock {\em ACM Transactions on Algorithms}, 3(2), 2007.

\bibitem{FKS84}
M.~L. Fredman, J.~Koml{\'o}s, and E.~Szemer{\'e}di.
\newblock Storing a sparse table with {$\Oh{1}$} worst case access time.
\newblock {\em Journal of the ACM}, 31(3):538--544, 1984.

\bibitem{GGKNP12}
T.~Gagie, P.~Gawrychowski, J.~K{\"a}rkk{\"a}inen, Y.~Nekrich, and S.~J.
  Puglisi.
\newblock A faster grammar-based self-index.
\newblock In {\em Proceedings of the 6th Conference on Language and Automata
  Theory and Applications (LATA)}, 2012.
\newblock To appear.

\bibitem{IR09}
C.~S. Iliopoulos and M.~S. Rahman.
\newblock Indexing factors with gaps.
\newblock {\em Algorithmica}, 55(1):60--70, 2009.

\bibitem{JMS04}
J.~J{\'a}J{\'a}, C.~W. Mortensen, and Q.~Shi.
\newblock Space-efficient and fast algorithms for multidimensional dominance
  reporting and counting.
\newblock In {\em Proceedings of the 15th International Symposium on Algorithms
  and Computation (ISAAC)}, pages 558--568, 2004.

\bibitem{KN11}
S.~Kreft and G.~Navarro.
\newblock Self-indexing based on {LZ77}.
\newblock In {\em Proceedings of the 22nd Symposium on Combinatorial Pattern
  Matching (CPM)}, pages 41--54, 2011.

\bibitem{MN07}
V.~M{\"a}kinen and G.~Navarro.
\newblock Rank and select revisited and extended.
\newblock {\em Theoretical Computer Science}, 387(3):332--347, 2007.

\bibitem{Mor68}
D.~R. Morrison.
\newblock {PATRICIA} - {Practical} algorithm to retrieve information coded in
  alphanumeric.
\newblock {\em Journal of the ACM}, 15(4):514--534, 1968.

\bibitem{Pat08}
M.~P\u{a}tra\c{s}cu.
\newblock Succincter.
\newblock In {\em Proceedings of the 49th Symposium on Foundations of Computer
  Science (FOCS)}, pages 305--313, 2008.

\bibitem{Tha11}
S.~V. Thankachan.
\newblock Compressed indexes for aligned pattern matching.
\newblock In {\em Proceedings of the 18th Symposium on String Processing and
  Information Retrieval (SPIRE)}, pages 410--419, 2011.

\bibitem{Wil83}
D.~E. Willard.
\newblock Log-logarithmic worst-case range queries are possible in space
  {$\Oh{N}$}.
\newblock {\em Information Processing Letters}, 7(2):81--84, 1983.

\end{thebibliography}

\end{document}